%% file: grundahl_rev.tex
\documentclass[mypaper,7pt,twoside]{CoAst}
\usepackage{epsf,graphicx,fancyhdr}
\input{CoAst_layo}

\def\note #1]{{\bf #1]}}

\begin{document}
\sf

\chapterDSSN{Stellar Oscillations Network Group}{}

\Authors{F. Grundahl$^1$, H. Kjeldsen$^1$, J. Christensen-Dalsgaard$^1$, 
T. Arentoft$^1$, S.~Frandsen$^1$} 
\Address{$^1$ Danish AsteroSeismology Centre (DASC), Department of Physics and Astronomy, University of Aarhus, Ny Munkegade, 8000 Aarhus C, Denmark}

\noindent
\begin{abstract}
Stellar Oscillations Network Group (SONG) is an initiative aimed at 
designing and building a network of 1m-class telescopes dedicated to 
asteroseismology and planet hunting. SONG will have 8 identical
telescope nodes each equipped with a high-resolution spectrograph and an 
iodine cell for obtaining precision radial velocities and a CCD 
camera for guiding and imaging purposes. The main asteroseismology targets
 for the network
are the brightest ($V\,<\,6$) stars. In order to improve performance and 
reduce maintenance costs the instrumentation will only have very few 
modes of operation. In this contribution we describe the motivations
for establishing a network, the basic outline of SONG and the expected 
performance.
\end{abstract}

\noindent
\section{Background and network motivation}

After the discovery of the global solar oscillations in the 1970's it 
was quickly realized that long continuous observations were needed in order
to obtain the best possible oscillation spectra. This ultimately led to the
construction of several networks, such as BiSON (Chaplin et al. 1996),
IRIS (Fossat 1991) and GONG (Harvey et al. 1996) dedicated to
the observation of the solar p-mode oscillations. 

In the study of oscillations in stars other than the Sun,
the limitations of short observing periods are well known, leading to 
aliasing problems in the observed power spectra resulting from 
a poor window function, and
low frequency precision caused by short observing runs. 

As was the case for the solar oscillations the best way to overcome this 
problem is to obtain long observing runs with high duty-cycle, and this 
demands either a groundbased telescope network or a space-based observatory
such as CoRoT or Kepler. 

During the past $\sim$5 years several teams have demonstrated the successful
detection of solar-like p-mode oscillations in other stars 
(Bedding et al. 2001, Bouchy et al. 2002) from time-series spectroscopy.
The development of methods
to measure high-precision velocities by groups hunting for extrasolar planets
has made the direct detection of solar-like oscillations in other stars 
possible. 

It is well known that the solar oscillations can be detected by measuring 
intensity variations or surface radial-velocity changes. In Fig. 1 we 
show the solar amplitude spectrum as measured 
in velocity (GOLF; Gabriel et al. 1995)
and intensity (VIRGO; Fr\"ohlich et al. 1995) by the SoHO satellite.
We note that the background is dramatically 
lower for the velocity signal compared with the intensity signal,
as already noted by Harvey (1988);
this demonstrates that velocity observations will be most efficient 
in detecting oscillations
in other stars. A further advantage of observing solar-like oscillations in
radial velocity is that modes with $l\,=\,3$ can be detected which is not
possible for intensity observations. 


\figureDSSN{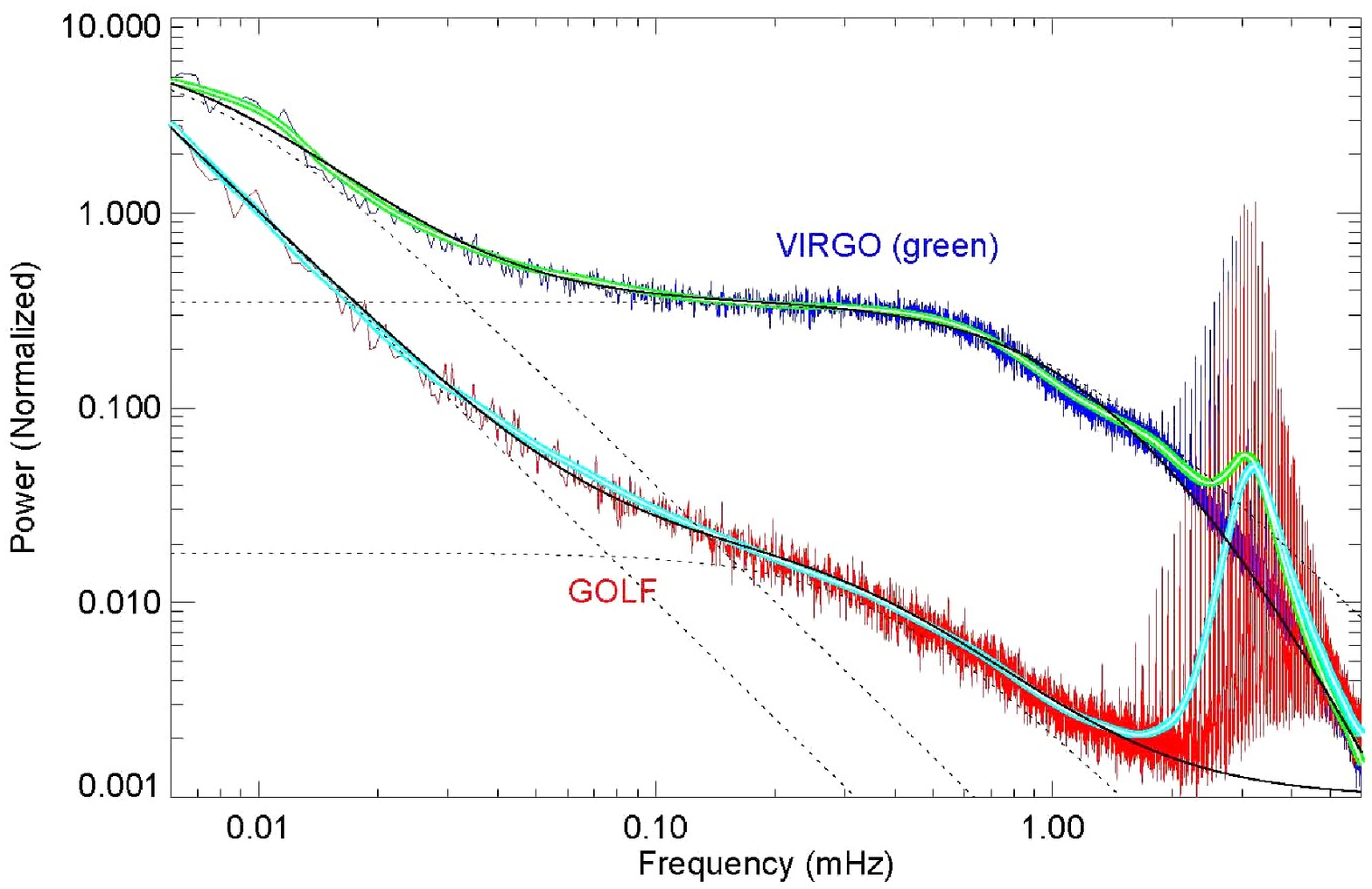}{
Comparison of data from VIRGO (green channel) and
GOLF. The power is normalized such that the p-mode amplitude for $l=1$
at peak power (near 3.1 mHz) is one for both VIRGO and GOLF. The background is
dominated by granulation and activity. A simple Harvey model is used to
describe the background (the different components shown as dashed curves).
The diagram also contains the smoothed power for both VIRGO and GOLF. At high
and low frequencies the p-mode signal-to-noise ratio (SNR)
is almost the same for GOLF and VIRGO. 
One should also note that the intensity background at frequencies above 
3--4 mHz is decreasing with frequency to the fourth power (which is not 
included in the Harvey model).
}{vel_vs_phot}{!ht}{width=100mm}

\noindent
\section{The need for a network}

As has been extensively discussed at this meeting asteroseismology has a 
great potential for increasing our understanding of stellar physics and 
evolution. The current instrumentation does not allow easy access to the 
facilities needed to provide long, un-interrupted velocity time series. At
the same time with the remarkable precision reached at the best instruments, 
such as HARPS (Mayor et al. 2003), UVES (Dekker et al. 2000), UCLES 
(Walker \& Diego 1985) and HiRES (Vogt et al. 1994)
it is also clear that the
access and availability of dedicated instrumentation is now 
the main limiting factor in the field. It is worth noting that the success of 
HARPS, UVES, UCLES and HiRES is due to the excellent quality of the 
instruments, more than a reflection of the primary mirror size. Thus a 
network dedicated to observing bright ($V\,<\,6$) stars will 
need high-quality instrumentation, more than aperture size -- 
this is a huge advantage
in terms of cost, since aperture is one of the main cost-drivers 
for large-aperture telescopes. 

A dedicated spectroscopic network will allow many different asteroseismic
projects to be carried out, including both long-term projects for a few
stars and short-term campaigns on several stars. Our
simulations show that it will be possible to determine reliably the large
and small frequency separations for solar-like stars in $\sim$one week of
observations, which for example could be used to determine the ages of
known planet-hosting stars and a significant fraction of the
DARWIN mission targets, typically FGK-type main-sequence stars.
On the other hand, observations over several months of a given star
will allow very detailed investigations of stellar internal properties,
utilizing also the expected reasonable SNR 
for even relatively low-order p modes whose frequency can be determined
with very high accuracy.
For many of the SONG targets it will
also be possible to determine radii from interferometric observations which
is a great help in the asteroseismic analysis. 

\noindent
\section{Network baseline}

To investigate whether a network such as SONG is realistic
a conceptual design study has been carried out during 2006 at the University
of Aarhus.
Here we briefly describe the current (autumn 2006) baseline 
for SONG. One of the main risks associated with the construction of a
network is the running costs and up-time of the instruments, and thus 
it is necessary to pay close attention to these issues. As a consequence 
of this we aim to limit the number of components in the dome to avoid 
exposure to ambient conditions and have as few moving parts a possible 
which implies a limited number of operation modes.  

The network will have 8 identical telescope nodes, four in each hemisphere, 
located at existing sites in order to avoid building significant new 
infrastructure. An illustration showing a possible location of sites is 
given in Fig. 2.  Each instrument will be remotely controlled -- for the 
long-term use of the network robotic observations are envisaged. It is,
however, an extremely complex task to robotise a telescope and 
hence full automatization may not
be achieved during the initial phases of operation. 

\figureDSSN{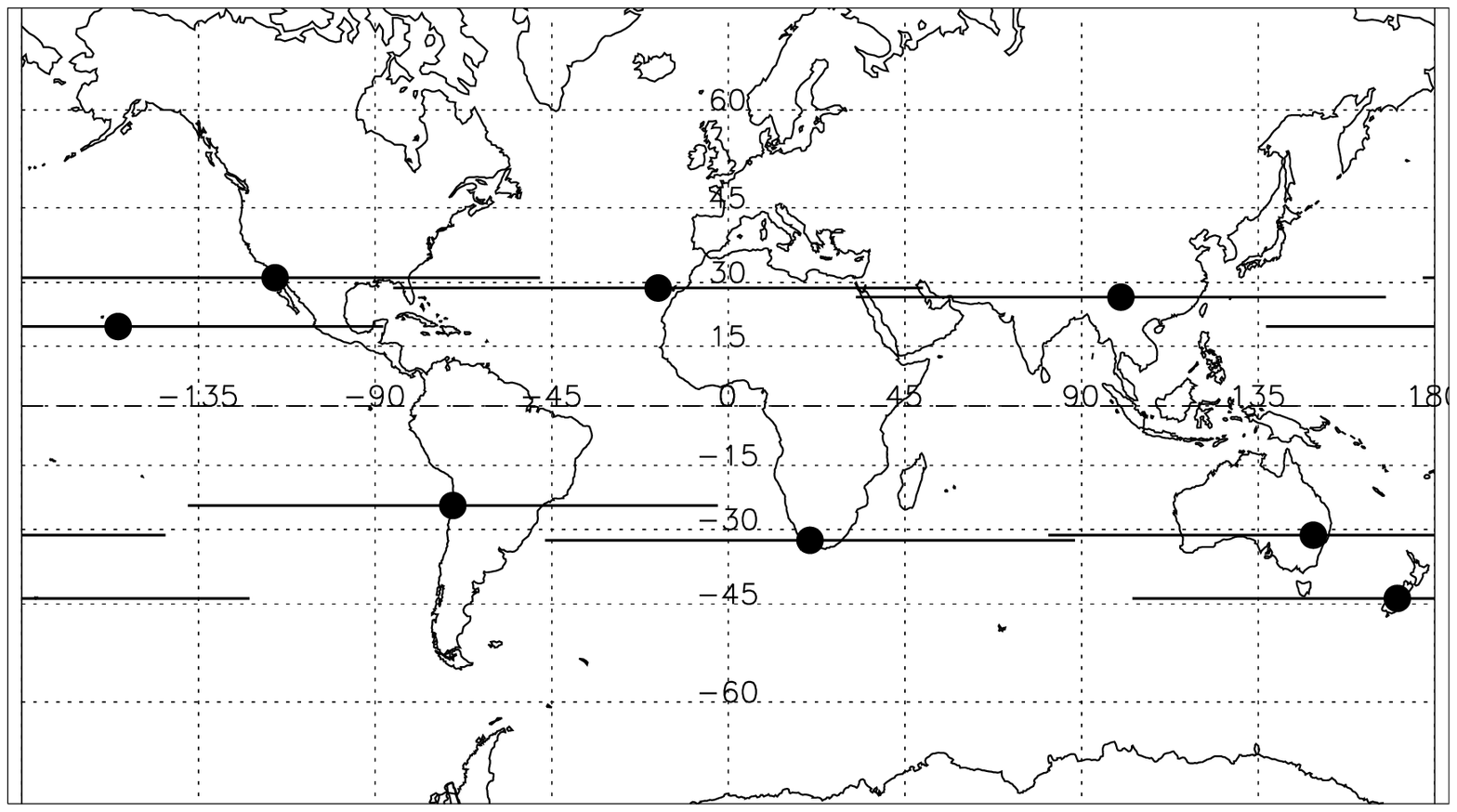}{A possible distribution of SONG sites, with
horizontal bars indicating the observability of an object which can be 
observed to $\pm4.5$ hours on either side of the meridian. For equatorial 
objects, which can be observed from both hemispheres, it would be possible
to obtain $\sim$60 hours of observation per 24 hours if all eight nodes
were observing the same object. Note that there will always be at least
two sites which can observe the same (equatorial) target, thus ensuring 
a high duty-cycle and valuable cross-checks on the measured velocities 
which will help to eliminate long-term drifts in the velocity zero points.
}{focal_plane}{!ht}{angle=00,width=100mm}

The conceptual design assumes
telescopes with a diameter of 80cm and an alt-az mount with a 
Coud{\'e} focus, housed in a dome with a diameter of 4m.
For the building we aim to use a standard 20 foot 
shipping container in which the two main instruments (spectrograph and 
imaging camera) will be located at the Coud{\'e} focus. The dome/building
configuration is similar in concept to that adopted by the Bradford Robotic
Telescope on Tenerife ({\tt http://www.telescope.org}). Our main motivation for 
choosing a Coud{\'e} focus is that this allows the dome to be completely
empty, apart from the telescope,
and to keep the instrumentation in a thermally controlled 
environment where all main components will be stationary -- this will  
be beneficial for reducing maintenance. 

Located at the focal station will be an optical table 
on which the instruments are mounted.
The main instrument will be a high-resolution spectrograph 
optimized for precision radial-velocity work. As velocity reference we 
will use an iodine cell in an arrangement similar to that developed by 
Butler et al. (1996). The spectrograph will be thermally isolated and
employ a UVES-like white-pupil design with a spectral resolution of $10^5$. 
An R4 echelle grating and a beam diameter of 75mm will result in a slit 
width of 1.5 arcsecond on the sky which will ensure a high throughput for most 
observing conditions. A 2K$\times$2K detector with low readout noise and
coatings optimized for the 500nm to 600nm region will be used to
record the spectrum -- this will possibly be a frame-transfer CCD which
would allow a very high duty-cycle. 
The spectral coverage will be
from 480nm to 670nm in order to cover the primary region of interest when 
using iodine and to also include the H$\alpha$ line. 
A preliminary optical design of the spectrograph
carried out at the Anglo Australian Observatory shows that essentially 
diffraction-limited image quality across the detector can be achieved with 
very little variation of the line-spread function. It is planned to include
also tip-tilt correction of the spectrograph feed in order to ensure maximum
throughput and reduce the effects of guiding and tracking errors. 
The spectrograph will have a fixed setup, 
although we may include a few slits of
fixed width to be able to change the spectral resolution. Figure 3 shows 
the basic outline of the telescope and focal plane.

\figureDSSN{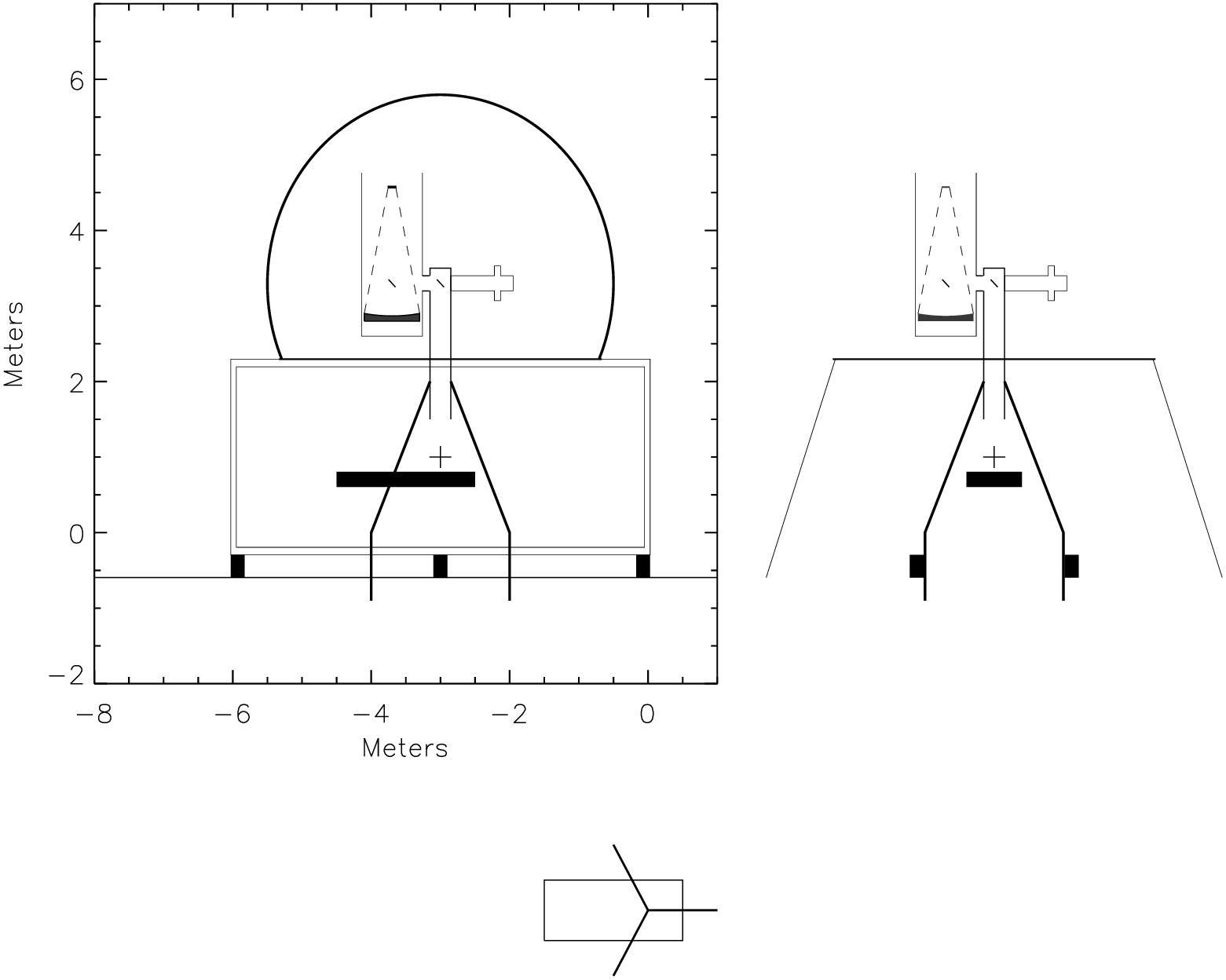}{Schematic layout of a SONG telescope and 
enclosure. The telescope focus is shown as a cross near $(x,y)=(-3,1)$ 
and the thick black bar is the optical table for the spectrograph and 
imager. The spectrograph and optical table will be thermally and 
mechanically  isolated from the surroundings. Note that this design 
only uses 4 mirrors. The configuration shown here is essentially a German 
equatorial mount with the polar axis in a vertical position -- this 
makes the design independent of the geographical latitude of the sites. 
The housing for the spectrograph is a standard 20 foot shipping container.
Such containers are very rugged and easily available. 
}{focal_plane}{!ht}{angle=0,width=100mm}

In front of the slit an atmospheric dispersion corrector (ADC) will be 
implemented, as well as calibration lamps and the temperature controlled 
iodine cell. 

Data are stored on-site for several weeks before being transported to a 
central institution; pipeline-reduced data will, however, be transmitted 
via the internet as soon as it has been processed by the data reduction 
pipeline. 


\section{Performance}

We have made a detailed assessment of the spectrograph performance based
on the AAO preliminary study and realistic numbers for seeing, slit width, 
mirror reflectivities and detector efficiency. The results are shown in 
Fig. 4 for a 75mm beam diameter spectrograph and a 1.25 arcsecond slit in 
2 arcsecond seeing at an airmass of two.

\figureDSSN{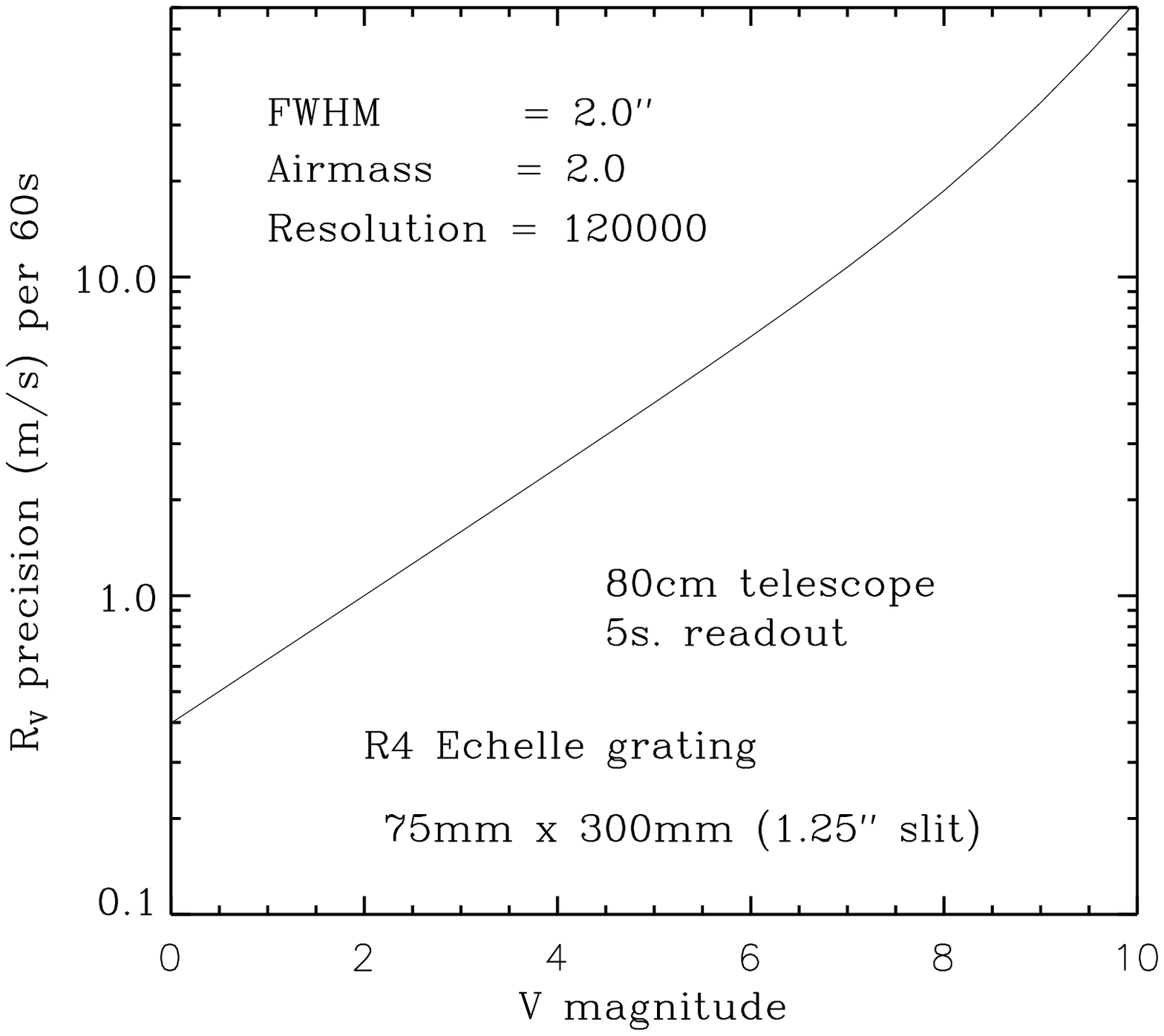}{
The predicted velocity precision for a single SONG node for a one minute 
observation versus the $V$ magnitude of the observed star. 
A spectral type similar to $\alpha$ Centauri A and a slow rotation has been
assumed. The echelle grating measures 75$\times$300mm, and the spectrograph 
has a collimated beam diameter of 75mm. The resolution with a 1.25 arcsecond
slit is around 120.000.
}{focal_plane}{!ht}{angle=00,width=100mm}

This performance compares well with that of UVES on VLT as 
reported by Butler et. al. (2004). The main reason that SONG performs
almost equally well as UVES on bright targets is the low duty-cycle
for UVES due to the long detector readout time, compared with the integration
time and to the narrow slit (0.3 arcsecond) needed to obtain the high
resolution. 

With this performance we have carried out simulations for solar-like stars
to see what would be required to estimate their ages based on the values
for their large and small frequency separations. The simulations show that
for stars brighter than $V\approx\,5$ these can be accurately determined 
from a one week observing campaign. 


\section{Status and schedule}

Currently (autumn 2006) SONG is nearly through its conceptual design 
phase. This is to be followed by detailed specifications and design of
all components for a prototype during 2007. We plan to have an extended
prototype phase (2008--2009) in order to eliminate all problems before going 
to full-scale operations, which is planned for around 2011--12. 

At {\tt http://astro.phys.au.dk/SONG} further
information and contact addresses for SONG can be found.

\acknowledgments{
The Carlsberg Foundation, the Villum Kann-Rasmussen foundation and the 
Danish Natural Science Research Council (FNU) are thanked for generous 
financial support to the conceptual design phase of this project.
}

\References{
Nature 439, 437 \\ 
Bedding, T. R., Butler, R. P., Kjeldsen, H. et al. 2001, ApJ 549, L105\\
Bouchy, F., Carrier, F. 2002, A\&A 390, 205\\
Butler, R. P., Marcy, G. W., Williams, E. et al. 1996, PASP 108, 500\\
Butler, R. P., Bedding, T. R., Kjeldsen, H. et al. 2004, ApJ 600, L75\\
Chaplin, W. J., Elsworth, Y., Howe, R. et al. 1996,
Solar Phys. 168, 1 \\ 
Dekker, H., D'Odorico, S., Kaufer, A. et al. 2000, Proc. SPIE 4008, 534\\
Fossat, E. 1991,
Solar Phys. 133, 1 \\ 
Fr\"ohlich, C., Romero, J., Roth, H. et al. 1995,
Solar Phys. 162, 101 \\ 
Gabriel, A. H., Grec, G., Charra, J. et al. 1995,
Solar Phys. 162, 61 \\ 
Harvey, J. W. 1988,
in Proc. IAU Symposium No 123,
eds J. Christensen-Dalsgaard \& S. Frandsen, 
Reidel, Dordrecht,
p. 497  \\ 
Harvey, J. W., Hill, F., Hubbard, R. P. et al. 1996,
Science 272, 1284 \\ 
Mayor, M., Pepe, F., Queloz, D. et al. 2003, The Messenger 114, 20\\
Vogt, S. S., Allen, S. L., Bigelow, B. C.  et al. 1994, Proc. SPIE 2198, 362\\
Walker, D. D., Diego, F. 1985, MNRAS 217, 355
}

\end{document}

%% file: CoAst_layo.tex
\renewcommand{\chaptermark}[1]{\markboth{#1}{}}

\newcommand{\kopf}{\small\itshape Comm. in Asteroseismology\\ Vol. 150, 2006}
\newcommand{\Authors}[1]{\begin{center}\normalsize\bf\sf #1 \end{center}}

\renewcommand{\author}[1]{\begin{center}\normalsize\bf\sf #1 \end{center}}
\newcommand{\Address}[1]{\begin{center}\small\sf #1 \end{center}}

\renewenvironment{abstract}{\section*{Abstract}\normalsize\sf}{}
\newcommand{\References}[1]{\begin{flushleft}{\large References\\}\vspace*{2mm}\small #1 \end{flushleft}}

\newcommand{\chapterDSSN}[2]{\chapter[\sf\normalsize #1\\ \footnotesize \hspace*{5mm}by #2 \sf\normalsize][]{#1\\}\rhead[\fancyplain{}{\sf\footnotesize \center{#1}}]{\fancyplain{}{\sffamily\thepage}}\lhead[\fancyplain{\kopf}{\sffamily\thepage}]{\fancyplain{\kopf}{\sf\footnotesize \center{#2}}}}

\newcommand{\figureDSSN}[5]{\begin{figure}[#4]
\centering
\includegraphics*[#5]{#1}
\caption{#2}
\label{#3}
\end{figure}}

\renewcommand{\chaptername}{}
\newcommand{\acknowledgments}[1]{\vspace*{5mm}\noindent\begin{bf}Acknowledgments. \end{bf} #1}